\documentclass[twocolumn]{aastex63}
\usepackage{amsmath}
\usepackage{booktabs}
\usepackage{xspace}
\usepackage{enumitem}

\received{January 15, 2021}
\revised{January 29, 2021}
\accepted{January 31, 2021}

\newcommand{\E}[1]{\ensuremath{\times 10^{#1}} }
\newcommand{\per}[1]{\rm\,#1\ensuremath{^{-1}}\xspace}

\newcommand\nicer{\textit{NICER}\xspace}
\newcommand\xmm{\textit{XMM-Newton}\xspace}
\newcommand\chandra{\textit{Chandra}\xspace}
\newcommand\integral{\textit{INTEGRAL}\xspace}
\newcommand\swift{\textit{Swift}\xspace}
\newcommand{\igr}{IGR J17494$-$3030}

\shorttitle{\igr{} X-ray Pulsations}
\shortauthors{Ng et al.}

\graphicspath{{./}{figures/}}

\begin{document}

\title{{\em NICER} Discovery of Millisecond X-ray Pulsations and an Ultracompact Orbit in \igr}

\correspondingauthor{Mason Ng}
\email{masonng@mit.edu}

\author[0000-0002-0940-6563]{Mason Ng}
\affiliation{MIT Kavli Institute for Astrophysics and Space Research, Massachusetts Institute of Technology, Cambridge, MA 02139, USA}

\author[0000-0002-5297-5278]{Paul S. Ray}
\affiliation{Space Science Division, U.S. Naval Research Laboratory, Washington, DC 20375, USA}

\author{Peter Bult}
\affiliation{Department of Astronomy, University of Maryland,
  College Park, MD 20742, USA}
\affiliation{Astrophysics Science Division, 
  NASA Goddard Space Flight Center, Greenbelt, MD 20771, USA}

\author[0000-0001-8804-8946]{Deepto Chakrabarty}
\affiliation{MIT Kavli Institute for Astrophysics and Space Research, Massachusetts Institute of Technology, Cambridge, MA 02139, USA}

\author[0000-0002-6789-2723]{Gaurava K. Jaisawal}
\affiliation{National Space Institute, Technical University of Denmark, Elektrovej 327-328, DK-2800 Lyngby, Denmark}

\author[0000-0002-0380-0041]{Christian Malacaria}
\affiliation{NASA Marshall Space Flight Center, NSSTC, 320 Sparkman Drive, Huntsville, AL 35805, USA}
\affiliation{Universities Space Research Association, NSSTC, 320 Sparkman Drive, Huntsville, AL 35805, USA}

\author[0000-0002-3422-0074]{Diego Altamirano} 
\affiliation{School of Physics and Astronomy, University of Southampton, Southampton, SO17 1BJ, UK}

\author{Zaven Arzoumanian}
\affiliation{Astrophysics Science Division, NASA Goddard Space Flight Center, Greenbelt, MD 20771, USA}

\author{Keith C. Gendreau}
\affiliation{Astrophysics Science Division, NASA Goddard Space Flight Center, Greenbelt, MD 20771, USA}

\author[0000-0002-3531-9842]{Tolga G\"{u}ver}
\affiliation{Istanbul University, Science Faculty, Department of Astronomy and Space Sciences, Beyazıt, 34119, Istanbul, Turkey}
\affiliation{Istanbul University Observatory Research and Application Center, Istanbul University 34119, Istanbul, Turkey}

\author[0000-0002-0893-4073]{Matthew Kerr}
\affiliation{Space Science Division, U.S. Naval Research Laboratory, Washington, DC 20375, USA}

\author[0000-0001-7681-5845]{Tod E. Strohmayer}
\affiliation{Astrophysics Science Division, NASA Goddard Space Flight Center, Greenbelt, MD 20771, USA}
\affil{Joint Space-Science Institute, NASA Goddard Space Flight Center, Greenbelt, MD 20771, USA}

\author[0000-0002-9249-0515]{Zorawar Wadiasingh}
\affiliation{Astrophysics Science Division, NASA Goddard Space Flight Center, Greenbelt, MD 20771, USA}
\affiliation{Centre for Space Research, North-West University, Potchefstroom Campus, Private Bag X6001, Potchefstroom 2520, South Africa}
\affiliation{Universities Space Research Association (USRA), Columbia, MD 21046, USA}

\author[0000-0002-5297-5278]{Michael T. Wolff}
\affiliation{Space Science Division, U.S. Naval Research Laboratory, Washington, DC 20375, USA}

\begin{abstract}
We report the detection of 376.05~Hz (2.66~ms) coherent X-ray pulsations in \nicer observations of a transient outburst of the low-mass X-ray binary IGR J17494$-$3030 in 2020 October/November. The system is an accreting millisecond X-ray pulsar in a 75~minute ultracompact binary. The mass donor is most likely a $\simeq 0.02\,M_\odot$ finite-entropy white dwarf composed of He or C/O. The fractional rms pulsed amplitude is 7.4\%, and the soft (1--3~keV) X-ray pulse profile contains a significant second harmonic. The pulsed amplitude and pulse phase lag (relative to our mean timing model) are energy-dependent, each having a local maximum at 4~keV and 1.5~keV, respectively. We also recovered the X-ray pulsations in archival 2012 {\em XMM-Newton} observations, allowing us to measure a long-term pulsar spin-down rate of $\dot\nu= -2.1(7)\E{-14}$~Hz~s$^{-1}$ and to infer a pulsar surface dipole magnetic field strength of $\simeq 10^9$~G. We show that the mass transfer in the binary is likely non-conservative, and we discuss various scenarios for mass loss from the system. 

\end{abstract}


\section{Introduction} \label{sec:intro}

Accreting millisecond X-ray pulsars \citep[AMXPs; see][for a recent review]{disalvo20} are rapidly rotating, weakly magnetized ($\sim 10^8$~G) neutron stars accreting from a low-mass ($\lesssim 1 M_\odot$) companion in a low-mass X-ray binary (LMXB). Most known AMXPs are X-ray transient systems in which long ($\sim$years) intervals of X-ray quiescence are punctuated by brief ($\sim$weeks) outbursts of enhanced X-ray emission.  These transient outbursts are understood to arise from a thermal instability in the accretion disk around a neutron star or black hole LMXB primary, analogous to ``dwarf nova'' optical outbursts in accreting white dwarfs \citep[see][and references therein]{Lasota2001,Hameury20}. 

The X-ray transient IGR J17494$-$3030 (Galactic coordinates $l=359.1^\circ$, $b= -1.5^\circ$; hereafter called IGR J17494) was first discovered in a 2012 March outburst in the 3--80~keV hard X-ray band (IBIS and JEM-X) in an {\em INTEGRAL} survey of the Galactic center region \citep{boissay12}.~Soft X-ray (0.5--10 keV) monitoring observations with \swift showed that the outburst lasted approximately one month \citep{armaspadilla2013} before fading into quiescence \citep{chakrabarty13}. \xmm 0.5--10~keV spectroscopy suggested that the compact  primary is a neutron star \citep{armaspadilla2013}. A new outburst was detected with \integral in 2020 October \citep{ducci20}, leading to a more precise X-ray localization with \chandra \citep{chakrabarty20} and the identification of a 4.5 GHz radio counterpart with the VLA \citep{vandeneijnden20}. 

Soft X-ray observations of the 2020 outburst with the {\em Neutron Star Interior Composition Explorer (NICER)} revealed the presence of coherent 376 Hz pulsations modulated by a 75~minute binary orbit, establishing the system as a millisecond pulsar (neutron star) in an ultracompact binary \citep{ngmason20}. In this Letter, we first outline the \nicer and \xmm observations and data processing. We then present results from timing and spectral analyses of the \nicer observations, as well as from a timing analysis of the archival 2012 \xmm observations. Finally, we constrain the possible nature of the donor in the IGR J17494 system and discuss further implications of the source.

\section{Observations and Data Processing} \label{sec:observations}
\subsection{\nicer} \label{sec:nicer_obs}

\nicer is an X-ray telescope mounted on the \textit{International Space Station} (\textit{ISS}) since 2017 June. \nicer has 56 aligned pairs of X-ray concentrator optics and silicon drift detectors (52 detectors are usually active on \nicer). \nicer is capable of fast-timing observations in the 0.2--12.0 keV band, with timing accuracy of time-tagged photons to better than 100 ns \citep{gendreau12,lamarr16,prigozhin16}.    

\nicer observed IGR J17494 from 2020 October 27 to November 4\footnote{The source became unobservable due to Sun-angle constraints around November~5.} for a total exposure time of $32.3{\rm\,ks}$ after filtering, in ObsIDs 3201850101--3201850108\footnote{During the course of the observations, several detectors were turned off for scheduled maintenance. Detectors 01, 02, 10, 13, 34, 43, and 44 were affected. In all observations, 46--48 detectors were active. Detectors 11, 20, 22, and 60 have been inactive since launch.}. These observations were available through the public NASA HEASARC data archive. There were additional \nicer observations, to which we did not have access, during this interval for a proprietary guest observer investigation (PI: A. Sanna; shown as the shaded region in the top panel of Figure \ref{fig:igrj17494general}). The events were barycenter-corrected in the ICRS reference frame, with source coordinates R.A. $= 267.348417\arcdeg$ and Decl.$ = -30.499722\arcdeg$ (equinox J2000.0) obtained from a recent \chandra observation \citep{chakrabarty20}, using \texttt{barycorr} from \texttt{FTOOLS} with the JPL DE405 solar system ephemeris \citep{standish98}. 

The \nicer observations were processed with \texttt{HEASoft} version 6.28 and the \nicer Data Analysis Software (\textsc{nicerdas}) version 7.0 (2020-04-23\_V007a). The following criteria, which we note are relaxed compared to standard filtering criteria as the latter were too restrictive and resulted in no events, were imposed in the construction of the good time intervals (GTIs): no discrimination of events when \nicer (on the \textit{ISS}) was inside or outside of the South Atlantic Anomaly during the course of the observations; $\geq 20\arcdeg$ for the source-Earth limb angle ($\geq 30\arcdeg$ for the Sun-illuminated Earth); $\geq$ 38 operational Focal Plane Modules (FPMs); undershoot (dark current) count-rate range of 0--400 per FPM (\texttt{underonly\_range}); overshoot (saturation from charged particles) count-rate range of 0--2 per FPM (\texttt{overonly\_range} and \texttt{overonly\_expr}); pointing offset is $<0.015\arcdeg$ from the nominal source position.

We analyzed spectral data using \texttt{XSPEC} v12.11.1 \citep{arnaud96}.~\nicer data were selected in the range 1--10~keV, to avoid contamination from optical loading and significant interstellar absorption at lower energy. The spectra were rebinned to have at least 25 counts per bin. Background spectra were extracted using \texttt{nibackgen3C50} version 6 from the official \nicer tools\footnote{\url{https://heasarc.gsfc.nasa.gov/docs/nicer/tools/nicer_bkg_est_tools.html}.}. Standard response files made available by the \nicer team were used to perform spectral analysis\footnote{\url{https://heasarc.gsfc.nasa.gov/docs/heasarc/caldb/data/nicer/xti/index.html}.}.

\subsection{\xmm} \label{sec:xmm_obs}
\xmm performed a $43{\rm\,ks}$ observation of IGR J17494 on
2012 March 31 (ObsID 0694040201). The \textsc{EPIC-PN} camera was operated in timing mode, yielding a time resolution of 29.56 $\mu$s, which is sufficient to allow us to search for the presence of coherent pulsations. We processed these data using SAS version 18.0 and the latest version of the calibration files\footnote{\url{https://www.cosmos.esa.int/web/xmm-newton/current-calibration-files}}. Applying standard screening criteria, we retained only those events with photon energies in the 0.4--10~keV range, with $\textsc{pattern}\leq4$ and screening $\textsc{flag} = 0$. Source events were extracted from \textsc{rawx} columns $[34:42]$, while background events were extracted from \textsc{rawx} $[51:59]$. Constructing a $32$-s resolution light curve of the source and background data, we find that the source count-rate gradually decreased over the span of the observation, dropping from 2 ct\per{s} to 1 ct\per{s}. Additionally, we filtered out an episode of background flaring that occurred between $15750{\rm\,s}$ and $21500{\rm\,s}$ after the start of the observation. Finally, we applied barycentric corrections to the cleaned event data, again using the JPL DE405 solar system ephemeris and the source coordinates quoted previously.

\section{Results} \label{sec:timing}
\subsection{NICER} \label{sec:nicertiming}

\begin{figure}[t]
	\centering
	\includegraphics[width=1.05\columnwidth]{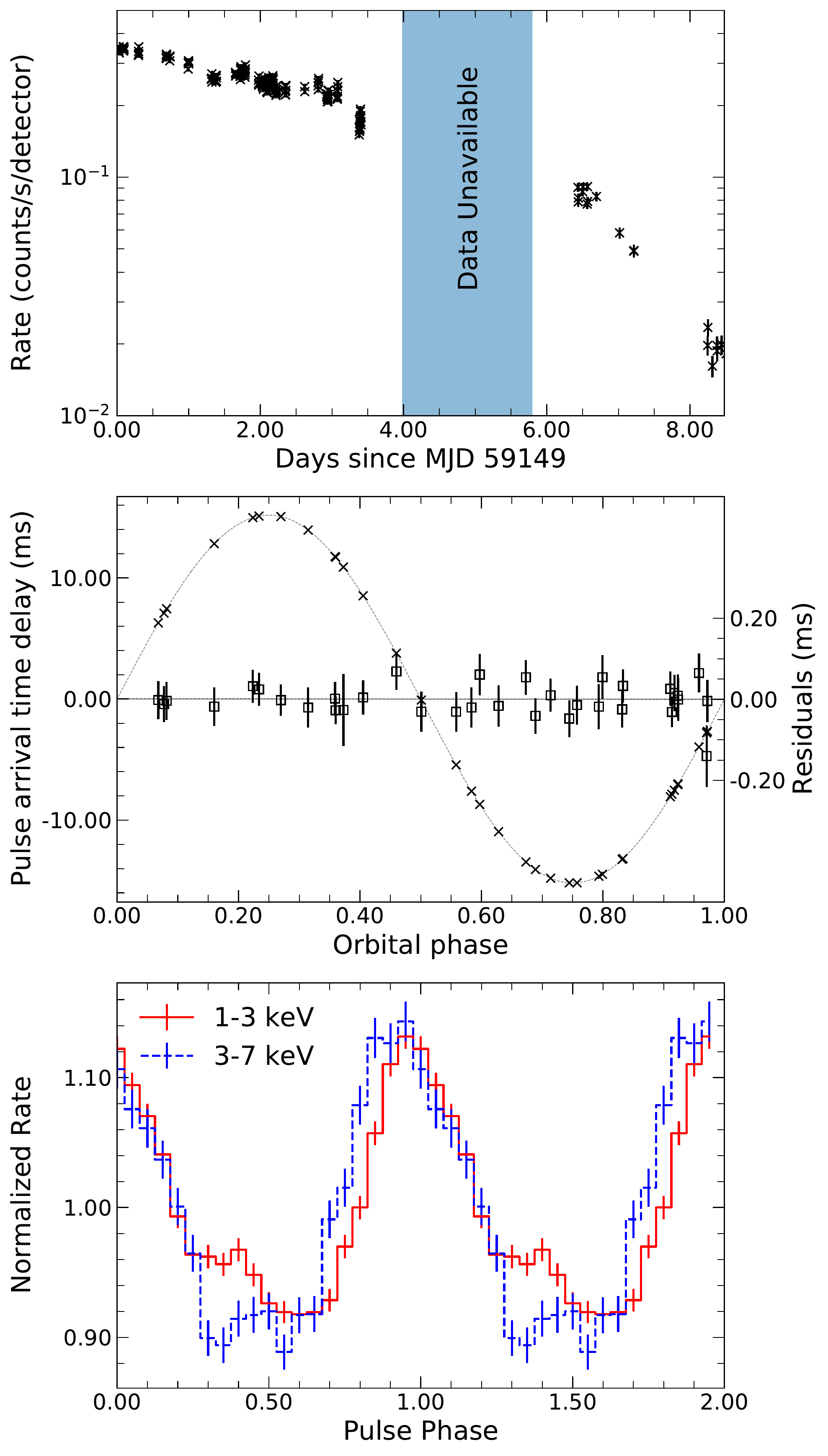}
	\caption{{\em Top:} \nicer 1--7~keV light curve for IGR J17494. The shaded band denotes a gap where proprietary \nicer data was unavailable to us.  
	{\em Middle:} Pulse arrival time delay as a function of orbital phase relative to the ascending node. The crosses are our measurements, and the solid curve is our best-fit model. The squares are the fit residuals, plotted on a 30$\times$ magnified scale. 
	{\em Bottom:} Pulse profiles in the 1--3~keV (solid red) and 3--7~keV (dashed blue) bands. The 1--3~keV profile contains a significant second harmonic.}
	\label{fig:igrj17494general}
\end{figure}

The \nicer 1--7~keV light curve for the 2020 outburst is shown in the top panel of Figure~\ref{fig:igrj17494general}. The source gradually faded until MJD 59155.4, after which it decayed more rapidly. The X-ray spectrum prior to the proprietary data gap was fairly constant and well-fit with a two-component absorbed power-law and blackbody model (\texttt{tbabs(powerlaw+bbodyrad)} in \texttt{XSPEC}), with absorption column density $n_{\rm H} = 2.07(6)\E{22}{\rm\,cm^{-2}}$, photon index $\Gamma = 1.90(6)$, blackbody temperature $kT = 0.58(3)$~keV, and blackbody radius $R_{\rm bb} = 2.9(5)\,d_{10}$~km, where $d_{10}$ is the source distance in units of 10~kpc. The uncertainties are reported at the 90\% confidence level. The reduced $\chi^2$ ($\chi_\nu^2$) of the fit was $1.14$ for 849 degrees of freedom. The spectrum softened during the late decay phase of the outburst, where the same two-component model fit yielded $\Gamma=4.3_{-0.6}^{+0.9}$ and $n_{\rm H}$ is assumed to be unchanged throughout the observations. The peak absorbed 1--10~keV flux we observed was $1.01\E{-10}{\rm\,erg\,s^{-1}\,cm^{-2}}$ on MJD $59149.4$, corresponding to an unabsorbed flux of $1.43\E{-10}{\rm\,erg\,s^{-1}\,cm^{-2}}$. The lowest absorbed flux we measured was $1.21\E{-12}{\rm\,erg\,s^{-1}\,cm^{-2}}$ on MJD $59157.5$, corresponding to an unabsorbed flux of $3.23\E{-12}{\rm\,erg\,s^{-1}\,cm^{-2}}$. This is roughly a factor of 3 fainter than the minimum flux detected by \xmm at the end of the 2012 outburst \citep{armaspadilla2013}. A more detailed X-ray spectral analysis will be reported elsewhere.

We first detected X-ray pulsations with a data analysis pipeline that employs multiple techniques\footnote{\url{https://github.com/masonng-astro/nicerpy\_xrayanalysis},~particularly with \texttt{Lv3\_incoming.py} and scripts therein} for X-ray pulsation searches, including averaged power spectral stacking with Bartlett's method \citep{bartlett48} and acceleration searches with \texttt{PRESTO} \citep{ransom02}.
The initial detection was made through \texttt{PRESTO}, an open-source pulsar timing software package\footnote{\url{https://github.com/scottransom/presto}} designed for efficient searches for binary millisecond pulsars. 
We ran a Fourier-domain acceleration search scheme with the \texttt{accelsearch} task over the range 1--1000~Hz, and posited that the Doppler motion would cause the possible signal to drift over a maximum of 100 bins in Fourier frequency space.
This yielded a strong $\simeq 376.05{\rm\,Hz}$ pulsation candidate (trial-adjusted significance of $3.5\sigma$) in the 2--12~keV range.

After initial identification of the candidate in the 2--12~keV range, we optimized the pulse significance by adjusting the energy range to maximize the $Z_1^2$ statistic, where 

\begin{equation} \label{eq:z1}
	Z_1^2 = \frac{2}{N}\left[\left( \sum_{j=1}^{N} \cos 2\pi\nu t_j \right)^2
	      + \left( \sum_{j=1}^{N} \sin 2\pi\nu t_j \right)^2\right],
\end{equation}
where $t_j$ are the $N$ photon arrival times \citep{Buccheri1983}. We found that an optimal energy range of 1.01--7.11~keV yielded $Z_1^2=1915.41$. Our subsequent timing analyses were carried out over 1--7~keV. 

\begin{figure}[t]
	\includegraphics[width=\linewidth]{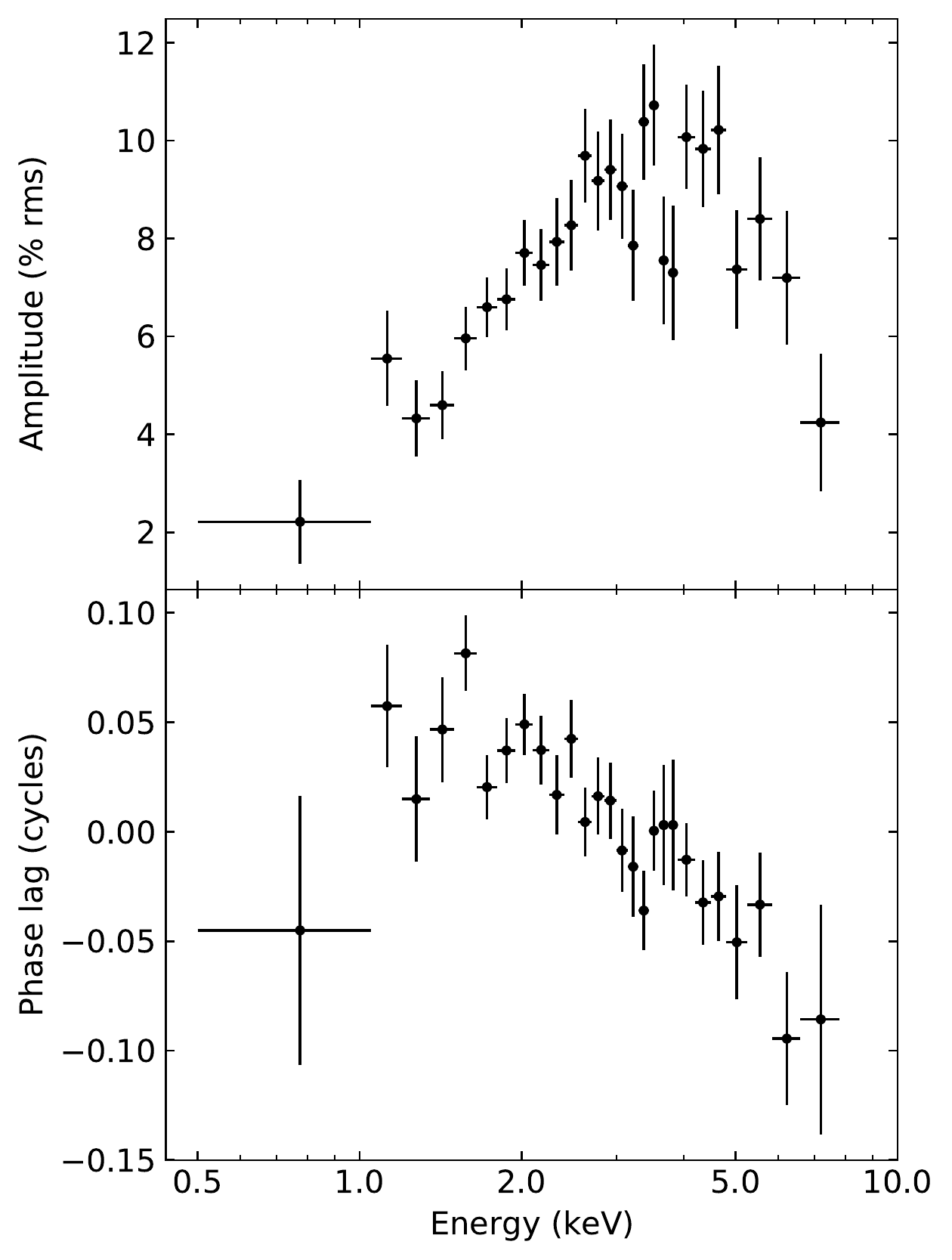}
	\caption{{\em Top:} Fractional rms pulsed amplitude as a function of energy, as measured by \nicer. {\em Bottom:} Pulse phase lag as a function of energy, as measured by \nicer. 
	The lag is measured relative to the best-fit timing model in Table~1.}
	\label{fig:pulse energy}
\end{figure}

The acceleration searches indicated that the pulsation frequency is modulated by a binary orbit. We used the acceleration data to estimate an initial timing model with a provisional circular orbit. We then used this initial model to construct $35$ pulse times of arrival (TOAs) with the \texttt{photon\_toa.py} tool in the \texttt{NICERsoft}\footnote{\url{https://github.com/paulray/NICERsoft}} data analysis package, using a Gaussian pulse template and ensuring an integration time of 500~s for each TOA (with minimum exposure time of 200~s). We then used these TOAs to compute corrections to our initial orbit model using weighted least-squares fitting with the \texttt{PINT} pulsar data analysis package \citep{luo20}. Our best-fit orbit ephemeris is shown in Table~\ref{tab:ephemeris}, and the orbital decay curve is shown in the middle panel of Figure~\ref{fig:igrj17494general}. Using our best-fit timing model, pulsations were detected throughout the entire outburst. At the end of the observations, we were able to detect the pulsations in observations from MJD 59154--59157 (November 1--4) by combining all the data. The mean unabsorbed flux over this 4-day interval was $8.5\E{-12}$ erg~s$^{-1}$~cm$^{-2}$ (1--10~keV). We did not have sufficient sensitivity to detect the pulsations in individual pointings from these dates. 
The time-averaged fractional root-mean-squared (rms) pulsed amplitude was 7.4\% (1--7~keV). Examining the lower and higher energies separately, we found amplitudes of 7.2\% in the 1--3~keV band and 8.7\% in the 3--7~keV band.  The soft and hard X-ray pulse profiles are shown in the bottom panel of Figure~\ref{fig:igrj17494general}. The 1--3~keV profile shows the presence of a second harmonic; this component is not significantly detected in the 3--7~keV profile.~To further examine the energy dependence of the pulse waveform, we adaptively binned the timing data in energy. We required the energy bins to contain a multiple of 5 pulse-invariant (PI) energy channels (0.05~keV), such that each bin contained at least 5000 counts. For each of these energy bins, we then folded the data using our best-fit timing solution and measured the background-corrected fractional rms pulsed amplitude and the pulse phase offset relative to the model. The resulting energy dependencies are shown in Figure \ref{fig:pulse energy}. The pulsed amplitude has a local maximum of 11\% at 4~keV, while the pulse phase lag has a local maximum of $+0.05$~cycles (130~$\mu$s) at around 1.5~keV. 

\subsection{XMM-Newton}

The uncertainty in our $P_{\rm orb}$ value does not allow us to coherently extrapolate our timing model back to the 2012 outburst. Thus, we searched for pulsations in the \xmm data by constructing a grid of trial $T_\mathrm{asc}$ values around the local epoch that spanned one orbital period. The grid resolution was set to 50~s, which is equivalent to $4\arcdeg$ in orbital longitude. For each trial ephemeris, we then demodulated the event data and computed the $Z_1^2$ statistic (see Eq.~\ref{eq:z1}). We evaluated this statistic for pulse frequencies in a $\pm3{\rm\,mHz}$ window around the spin frequency measured with \nicer, adopting a frequency resolution of $1/T$, with $T$ the duration of the \xmm observation. The best candidate solution produced by this search had $Z_1^2 = 89$, which converts to a trial-adjusted pulse detection significance of $8\sigma$. 

Adopting the best $T_\mathrm{asc}$ and pulse frequency from the grid search as a provisional model, we performed a phase-coherent pulse analysis.~We divided the light curve into $\approx3$\,ks segments, and measured the pulse phase in each segment separately. The phase residuals were fit using a circular orbital model and constant spin frequency, where we kept the orbital period and projected semimajor axis fixed at their \nicer values. The best-fit values were $\nu_{2012}=376.0501759(19)$~Hz and $T_{\rm asc,2012}=$ MJD $56017.33680(5)$.  Comparing to our \nicer measurement, we find $\Delta\nu \equiv  \nu_{2020} - \nu_{2012} = -5.7\pm1.9$~mHz. This indicates long-term spin-down of the pulsar between outbursts, at a rate $\dot\nu = -2.1(7)\E{-14}$\,Hz\,\per{s}. Owing to the uncertainty in exact orbital cycle count between the 2012 and 2020 epochs, we are unable to use these $T_{\rm asc}$ measurements to further refine the orbital period.

The \xmm data also showed an energy-dependent trend in pulse phase lag similar to that observed in the \nicer data.  We were unable to measure an energy-dependence in the pulsed amplitude with \xmm, but the results from the two data sets were consistent within the measurement uncertainties. 

\begin{table*}[t]
\centering
\caption{
  \igr{} timing parameters from the 2020 outburst
  \label{tab:ephemeris}
}  
\begin{tabular}{lc}
\toprule 
\toprule
Parameter & Value \\
\toprule 
Right ascension, $\alpha$ (J2000) & $267.348417\arcdeg$ \\
Declination, $\delta$ (J2000) & $-30.499722\arcdeg$ \\
Position epoch (TT) & MJD $59156.34$ \\
Spin frequency, $\nu_0$ (Hz) & $376.05017022(4)$ \\
Spin frequency derivative (during outburst), $|\dot{\nu}|$ (Hz/s) & $ < 1.8\E{-12}$ \\
Spin epoch, $t_0$ (TDB) & MJD $59149.0$ \\ 
Binary period, $P_{\rm orb}$ (s) & $4496.67(3)$ \\
Projected semimajor axis, $a_x \sin i$ (lt-ms) & $15.186(12)$ \\
Epoch of ascending node passage, $T_{\rm asc}$ (TDB) & MJD $59149.069012(15)$ \\ 
Eccentricity, $e$ & $<0.006\ (2\sigma)$ \\
\midrule 
Spin frequency derivative (long-term), $\dot{\nu}$ (Hz/s) & $-2.1(7)\times10^{-14}$ \\
\bottomrule
\end{tabular}

\tablecomments{Source coordinates adopted for the barycentering were determined by \cite{chakrabarty20}. The spin frequency derivative quoted here is during the 2020 outburst.}
\end{table*}

\section{Discussion} \label{sec:discussion}

The discovery of coherent millisecond X-ray pulsations from \igr{} definitively identifies the source as an accreting neutron star.  We can use the long-term spin-down of the pulsar between its 2012 and 2020 X-ray outbursts to estimate the pulsar's magnetic field strength. Assuming that the spin-down is due to magnetic dipole radiation, we can calculate the pulsar's magnetic dipole moment \citep{Spitkovsky2006}

\begin{align}
    \mu &= 5.2\E{26}
	\left( 1 + \sin^2 \alpha \right)^{-1/2} 
	\nonumber \\ &\times
	\left( \frac{I}{10^{45} \text{ g cm}^2} \right)^{1/2}
	\text{G cm}^3,
\end{align}
where $\alpha$ is the angle between the magnetic and spin axes, and $I$ is the neutron star moment of inertia. The corresponding surface dipole field strength of $\simeq 10^9$~G is on the high end of the distribution inferred for other AMXPs \citep{Mukherjee2015}.

We found that the fractional rms pulsed amplitude and the pulse phase of IGR J17494 vary as a function of photon energy. Both the amplitude and the phase lag reach a local maximum at a (different) characteristic energy of 4 and 1.5~keV, respectively. Energy-dependent variations of the pulse waveform are ubiquitous among AMXPs, although the location of these local maxima varies greatly from source to source \citep{Gierlinski2002, Gierlinski2005, Falanga2005b, Patruno2009d, Falanga2012}. The behavior can be understood through a two-component emission model, with thermal emission originating from the stellar surface and scattered Compton emission originating from some height above the surface \citep{Gierlinski2002, wilkinson11}. Accounting for the difference in geometry and emission patterns, such a model can self-consistently explain the energy dependence of both the phase lags and the pulsed amplitudes \citep{Poutanen2003}. 

Our measurement of a 75~min binary orbit allows us to constrain the nature of the mass donor in this system. The vast majority of Roche-lobe--filling LMXBs and cataclysmic variables contain hydrogen-rich donor stars, and they all have binary periods $P_{\rm orb}\gtrsim 80$ min \citep{paczynski81b,rappaport82}. The so-called ultracompact binaries ($P_{\rm orb}\lesssim 80$~min) have H-depleted donors \citep{Nelson1986,Pylyser1988,Pylyser1989,Nelemans2010}. IGR J17494 has the longest known period for an ultracompact LMXB and lies near the period boundary, making it a particularly interesting case. We also note the recent discovery of the rotation-powered millisecond gamma-ray pulsar PSR J1653$-$0158 in a 75~min (non-accreting) binary \citep{Nieder2020}. This is the {\em shortest} orbital period known for a rotation-powered binary pulsar, and this ``black widow'' system is believed to have evolved from an ultracompact LMXB after mass transfer ended. 

\begin{figure}[t]
	\centering
	\includegraphics[width=\columnwidth]{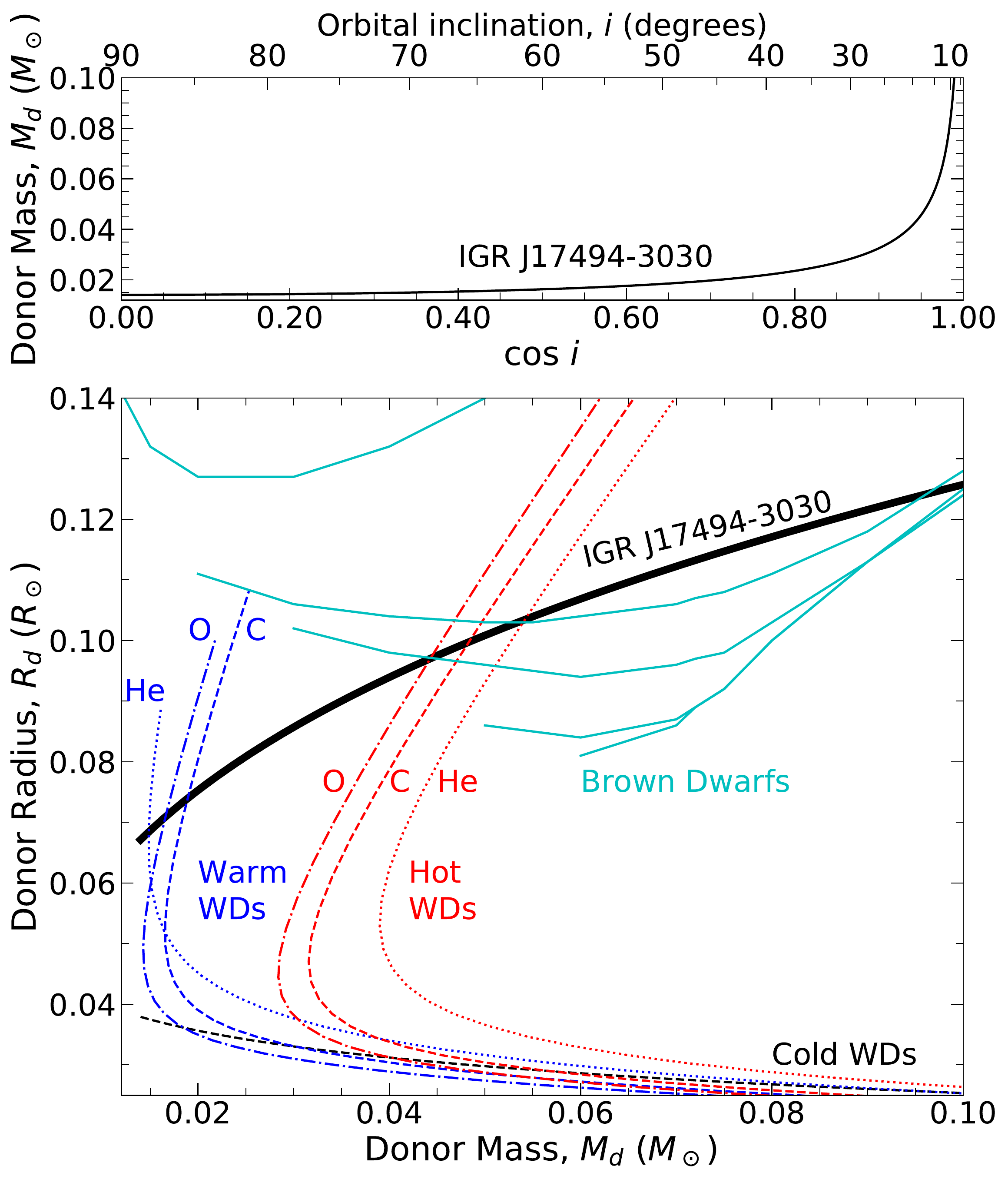}
	\caption{{\em Top:} Donor star mass $M_d$ as a function of binary inclination $i$, assuming $M_{\rm ns}=1.4\, M_\odot$. The a priori probability distribution is uniform in $\cos i$, so low masses are likeliest. {\em Bottom:} Mass-radius constraints for the donor star. The thick solid black curve is the mass-radius constraint for a Roche-lobe--filling donor from our orbital measurements. The dashed black line shows cold WD models. The blue and red lines show representative ``warm'' and hot WD models, respectively, with He (dotted), C (dashed), and O (dash-dotted) compositions. These models take $T_c=2.5$ and 7.9~MK for He and $T_c=5$ and 10~MK for C/O. The solid cyan curves show brown dwarf models for ages 0.1, 0.5, 1.0, 5.0, and 10.0 Gyr (from top to bottom). The likeliest donor is a warm $\simeq 0.02 M_\odot$ He or C/O WD.}
	\label{fig:massradius}
\end{figure}

From our measured orbital parameters, the binary mass function of IGR J17494 is 

\begin{align} \label{eq:massfunc}
    f_m & \equiv \frac{(M_d \sin i)^3}{(M_{\rm ns} + M_d)^2} = \frac{4\pi^2 (a_x \sin i)^3}{G P_{\rm orb}^2} \nonumber \\
      &\approx 1.39\E{-6}\ M_\odot, \end{align}
where $M_{\rm ns}$ is the neutron star mass, $M_d$ is the donor mass, $a_x \sin i$ is the projected semimajor axis, and the binary inclination $i$ is defined as the angle between the line of sight and the orbital angular momentum vector. For a given value of $M_{\rm ns}$, we can use Equation~\ref{eq:massfunc} to calculate $M_d$ as a function of $i$ (see top panel of Figure~\ref{fig:massradius}). Assuming $M_{\rm ns}=1.4\ (2.0)\, M_\odot$, the minimum donor mass (for an edge-on binary with $i=90^\circ$) is 0.014 (0.018) $M_\odot$. For a random ensemble of binaries, the probability distribution of $\cos i$ is uniformly distributed and $\Pr(i<i_0) = 1 - \cos i_0$. Thus, the donor mass is likely to be very low, with a 90\% confidence upper limit of $M_d < 0.033\ (0.041)\, M_\odot$ for $M_{\rm ns}=1.4\ (2.0)\, M_\odot$. 

Assuming a Roche-lobe--filling donor, we can calculate the donor radius $R_d$ as a function of $M_d$ \citep{eggleton83}; this is shown in the bottom panel of Figure~\ref{fig:massradius} for $M_{\rm ns}=1.4\, M_\odot$. For comparison, the figure also shows the mass-radius relations for different types of low-mass stars: cold white dwarfs \citep[WDs;][]{zapolsky69,rappaport84,nelemans01}; hot (finite-entropy) WDs composed of either He, C, or O \citep{deloye03}; and low-mass H-rich stars, including brown dwarfs \citep{chabrier00}. We see that cold WD models are inconsistent with our measured mass-radius constraint, indicating that thermal bloating is likely important. Moderately hot He WDs with central temperature $T_c=2.5\times 10^6$~K or C/O WDs with $T_c=5\times 10^6$~K are consistent with our constraint at high binary inclination. Hotter WDs and moderately old (cool) brown dwarfs are also consistent, but the required inclinations have low a priori probability.~Finally, H-rich dwarfs above the mass-burning limit are also possible, but only for extremely low (improbable) inclinations. We conclude that the donor is likely to be a $\simeq 0.02\, M_\odot$ finite-entropy He or C/O white dwarf. 

The angular momentum evolution of the binary is described by \citep{verbunt1993,verbunt95}

\begin{equation} \label{eq:angmom}
    -\frac{\dot{J}}{J} = -\frac{\dot{M}_d}{M_d}\,f_{\rm ML},
\end{equation}
where $\dot{J}$ is the rate of change of the orbital angular momentum $J$ due to effects other than mass loss from the system, $\dot{M}_d$ ($<0$) is the rate of change of the donor mass, and the dimensionless factor $f_{\rm ML}$ is given by

\begin{equation}
f_{\rm ML} = \frac{5}{6}+\frac{n}{2}-\beta q - \frac{(1-\beta)(q+3\alpha)}{3(1+q)} ,
\label{eq:Fmr}
\end{equation}
where $q=M_d/M_{\rm ns}\ll 1$ is the binary mass ratio, $\beta$ is the fraction of $\dot M_d$ that accretes onto the neutron star ($\beta=1$ for conservative mass transfer),

\begin{equation}
n = \frac{d (\ln R_d)}{d (\ln M_d)}
\end{equation}
denotes how the donor radius $R_d$ changes with mass loss, and $\alpha$ is the specific angular momentum of any (non-conservative) mass lost from the system in units of the donor star's specific angular momentum. Thus, $\alpha$ parameterizes the site of any mass ejection from the system, where $\alpha \simeq 1$ for mass loss close to the donor and $\alpha \simeq q^2$ for mass loss close to the pulsar.
Mass transfer in ultracompact binaries is primarily driven by angular momentum loss due to gravitational radiation from the binary orbit \citep[see][and references therein]{rappaport82}; for a circular orbit, this loss is given by \citep{landau89,peters64}

\begin{equation}
-\left(\frac{\dot J}{J}\right)_{\rm GW} = 
\frac{32\, G^3}{5\, c^5} \frac{M_{\rm ns}M_d (M_{\rm ns}+M_d)}{a^4} ,
\end{equation}
where $a$ is the binary separation. Inserting this into the left-hand side of Equation~\ref{eq:angmom}, we can then calculate the gravitational-wave--driven mass transfer rate from the donor into the accretion disk as 

\begin{equation*} \dot{M}_{\rm GW} = -\dot{M}_d = \frac{32 G^3}{5c^5} \left(\frac{4\pi^2}{G}\right)^{4/3} \frac{M_{\rm ns}^{8/3}\,q^2}{(1+q)^{1/3}\,P_{\rm orb}^{8/3}\,f_{\rm ML}}
\end{equation*}
\begin{equation*} \approx 2.6\E{-12}  \left(\frac{M_{\rm ns}}{1.4M_\odot}\right)^{2/3}
\end{equation*}
\begin{equation} \label{eq:mdot}
\times \left(\frac{M_d}{0.014M_\odot}\right)^2 \left(\frac{f_{\rm ML}}{0.66}\right)^{-1} M_\odot \mbox{\rm\, yr$^{-1}$}.
\end{equation}
Our scaling value of $f_{\rm ML}=0.66$ corresponds to $n=-1/3$ (typical for degenerate donors) and $\beta=1$.

Although accretion onto the neutron star is mediated by episodic outbursts, mass continuity requires that the long-term average accretion luminosity reflect $\dot M_{\rm GW}$ if the mass transfer is conservative.  Our observations are not ideal for examining this, since we did not observe the early (brightest) part of the 2020 outburst with {\em NICER}. However, the unabsorbed 0.5--10~keV X-ray fluence in the 2012 outburst was $1.1\E{-4}$ erg\,cm$^{-2}$ \citep{armaspadilla2013}. Assuming that the 2012 outburst was typical, that the long-term average accretion rate is dominated by the outbursts, and that there were no intervening outbursts between 2012 and 2020, the outburst separation of $\approx 3100$~days yields a long-term average X-ray flux of $F_{x,{\rm avg}} = 3.9\E{-13}$ erg s$^{-1}$ cm$^{-2}$ (0.5--10~keV). We can then write the accretion luminosity as
 
\begin{equation} \label{eq:acclum}
\frac{G M_{\rm ns}\beta\dot{M}_{\rm GW}}{R_{\rm ns}} = \left(\frac{\Delta\Omega}{4\pi}\right) 4\pi d^2 f_{\rm bol}\, F_{x,{\rm avg}}, 
\end{equation} 
where $R_{\rm ns}$ is the neutron star radius, $d$ is the distance to the source, $f_{\rm bol}$ is the bolometric correction (accounting for accretion luminosity outside the 0.5--10~keV bandpass), and $\Delta\Omega$ is the solid angle into which the accretion luminosity is emitted. Based on the {\em INTEGRAL} hard X-ray observations in 2012 \citep{boissay12}, we estimate $f_{\rm bol}\approx 1.7$. Assuming $R_{\rm ns}=10$~km and taking $\beta=1$ and $\Delta\Omega=4\pi$, we obtain an implausibly large distance of 20~kpc. Although it is not impossible that the source lies on the far side of the Galaxy, a location near the Galactic center is far more likely given the line of sight. There are several reasons that our distance estimate might be significantly inflated. Obtaining a more plausible distance of 8~kpc would require

\begin{eqnarray}
\frac{1}{\beta}
\left(\frac{\Delta\Omega}{4\pi}\right)
\left(\frac{f_{\rm bol}}{1.7}\right)
\left(\frac{f_{\rm ML}}{0.66}\right) 
& & \nonumber \\
\times \left(\frac{M_{\rm ns}}{1.4\,M_\odot}\right)^{-5/3}
\left(\frac{M_d}{0.014\,M_\odot}\right)^{-2} 
& & \nonumber \\
\times \left(\frac{F_{x,{\rm avg}}}{3.9\E{-13} \mbox{\rm\ erg~s$^{-1}$~cm$^{-2}$}}\right)
& \approx & 6 .
\end{eqnarray}
Some combination of these factors may be different than what we assumed above. However, a heavier neutron star ($M_{\rm ns}>1.4\,M_\odot$), a heavier mass donor (equivalent to a lower binary inclination), or significant beaming ($\Delta\Omega < 4\pi$) would further inflate the distance estimate. Also, our estimate of $f_{\rm bol}$ is fairly robust, given the broad X-ray coverage of the {\em INTEGRAL} data. It is possible that we have underestimated $F_{x,{\rm avg}}$. This could happen if we missed accretion outbursts that occurred between 2012 and 2020, or if the quiescent (non-outburst) flux is as high as $\sim 10^{-12}$~erg~s$^{-1}$~cm$^{-2}$.  The former possibility can be explored through a careful analysis of archival X-ray monitoring data, while the latter possibility could be checked through sensitive X-ray observations of the source in quiescence.  

The factor $f_{\rm ML}$ may be somewhat larger than we assumed.  Although we calculated it using the usual value of $n= -1/3$ for degenerate donors, \citet{deloye03} showed that the WD donors in ultracompact binaries can have $n$ values in the range of $-0.1$ to $-0.2$ due to the importance of Coulomb interactions for extremely low donor masses. However, this is unlikely to increase $f_{\rm ML}$ by more than a factor of $\simeq 1.2$. 

Non-conservative mass transfer ($\beta<1$) is a more promising avenue. The radio detection of IGR J17494 \citep{vandeneijnden20} points to the likelihood of a collimated jet ejection during the outburst. Moreover, a similar distance conundrum was invoked to infer non-conservative mass transfer in the ultracompact LMXB pulsar XTE J0929$-$314 \citep{marino17} as well as several other AMXPs \citep{marino19}. Also, there was evidence found for an outflow in the ultracompact LMXB pulsar IGR J17062$-$6143 \citep{degenaar17,vandeneijnden18}, possibly arising from a magnetic propeller-driven wind from the inner accretion disk \citep{illarionov75}.

During the long periods of X-ray (accretion) quiescence, mass loss from the binary could arise from several different mechanisms. These are motivated by the study of rotation-powered radio millisecond pulsars in detached (non-accreting) binaries: the so-called ``black widow'' ($M_c \lesssim 0.05 M_\odot$) and ``redback'' ($M_c \gtrsim 0.1 M_\odot$) systems, where $M_c$ is the companion mass \citep[see, e.g.,][and references therein]{romani16}. One possibility is black-widow--like ablation of the companion, driven by rotation-powered gamma-ray emission from the pulsar \citep{ginzburg20}.  Such ablation could also be driven by particle heating via the rotation-powered pulsar wind \citep[see][and references therein]{harding90}. Hard X-rays and gamma-rays from the intrabinary shock observed in many black widow systems could significantly affect the mass loss rate \citep{wadiasingh18}. Another possibility is that the pulsar wind could drive an outflow from the inner Lagrange ($L_1$) point by overcoming the ram pressure of accreting material \citep{burderi01,disalvo08}. 

 As an example, we consider the case of gamma-ray ablation. If we assume the gamma-ray luminosity is $\simeq10\%$ of the spin-down luminosity ($\simeq 3\E{35}$ erg~\per{s} based on our long-term $\dot\nu$ measurement) as typically seen in black widow systems \citep{abdo13}, this would imply a companion mass loss rate of $\sim10^{-11}M_\odot/{\rm yr}$ \citep{ginzburg20}. For a source distance of 8~kpc and assuming that gravitational wave losses dominate in Equation~\ref{eq:angmom}, this implies $\beta\approx0.04$ and $\alpha\approx0.4$, suggesting that the mass ejection occurs somewhere between the pulsar and the $L_1$ point ($\alpha\approx0.8$). However, \citet{ginzburg20} argue that magnetic braking of the donor (through magnetic coupling to the ablated wind) likely dominates gravitational radiation as an angular momentum sink in black widow systems. If so, then that could both decrease $\beta$ and increase $\alpha$ even further in our case.
 
 All of the X-ray--quiescent mechanisms mentioned above rely on the system entering a rotation-powered radio pulsar state during X-ray quiescence.  We note that a growing class of so-called transitional millisecond pulsars (tMSPs) has been identified that switch between LMXB and radio pulsar states \citep[see][for a review]{papitto20}. The known tMSPs would be classified as redback systems in their radio pulsar state. If IGR J17494 is a tMSP, then its low companion mass would make it a black widow system in its rotation-powered state. We note that the X-ray properties of IGR J17494 correspond to those of the so-called very faint X-ray transients \citep[VFXTs;][]{wijnands06}, whose low outburst luminosities and long-term accretion rates are difficult to understand.  Our observations support the suggestion that some VFXTs may also be tMSPs \citep{heinke15}. The distinction between VFXTs and ordinary LMXBs may somehow relate to the level of non-conservative mass transfer. 

\acknowledgments M.N. and this work were supported by NASA under grant 80NSSC19K1287 as well as through the \nicer mission and the  Astrophysics Explorers Program. \nicer work at NRL is also supported by NASA. D.A. acknowledges support from the Royal Society. This research has made use of data and/or software provided by the High Energy Astrophysics Science Archive Research  Center (HEASARC), which is a service of the Astrophysics Science Division at NASA/GSFC and the High Energy Astrophysics Division of the Smithsonian Astrophysical Observatory.

\facilities{NICER, XMM}

\software{astropy \citep{astropy}, NumPy and SciPy \citep{oliphant07}, Matplotlib \citep{hunter07}, IPython \citep{perez07}, tqdm \citep{dacostaluis20}, NICERsoft, PRESTO \citep{ransom02}, PINT \citep{luo20}, HEASoft 6.28}

\bibliography{mason}
\bibliographystyle{aasjournal}

\end{document}